\documentclass[ reprint,
 amsmath,amssymb,
 aps,
]{revtex4-2}
\usepackage{ulem}
\newcommand\redout{\bgroup\markoverwith
{\textcolor{red}{\rule[.5ex]{2pt}{0.4pt}}}\ULon}
\usepackage{xcolor}
\usepackage{physics}
\usepackage{mathrsfs}
\usepackage{graphicx}
\usepackage{dcolumn}
\usepackage{bm}
\usepackage{hyperref}
\usepackage[mathlines]{lineno}

\begin{document}

\preprint{APS/123-QED}

\title{On the Generation of  Topological Vector Solitons from Bessel Like Beams}
\author{Finn Buldt$^1$} \author{Pascal Bass\`ene$^1$} \author{Moussa N'Gom$^{1,2,}$}
\thanks{Corresponding Author: ngomm@rpi.edu}%
\affiliation{$^1$Department of Physics, Applied Physics and Astronomy, Rensselaer Polytechnic Institute, Troy, NY 12180, USA} 
\affiliation{$^2$Center for Ultrafast Optical Science, University of Michigan, Ann Arbor, MI 48109, USA} 



\begin{abstract}
Coupled solitary waves in optics literature, are coined vector solitons to reflect their particle--like nature that remains intact even after mutual collisions. They are born from a nonlinear change in the refractive index of an optical material induced by the light intensity.
We've discovered that the second harmonic intensity profile generated by  Bessel-like beams, is composed of solitons of various geometries surrounded by concentric rings; one of which is two central solitons of similar radius knotted by ellipsoidal concentric rings. We observe that their geometry and intensity distribution is dependent on the topological charge of the fundamental Bessel beams incident on the nonlinear medium. We show that their spatial profile is invariant against propagation. We observe that their behavior is similar to  that of  screw dislocations in wave trains: they collide and rebound at a $90^\circ$ angle from the beam propagation direction. In this way, we have generated linked frequency doubled Bessel-type vector solitons with different topologies, that are knotted as they oscillate along the optical axis, when propagating in the laboratory environment.
\end{abstract}

\maketitle
\section{Introduction}
Optical solitons  are unusually stable solitary waves. They  retain their shape as they propagate over long distances  without any noticeable spread in width or loss of amplitude.  Each soliton pulse remains localized within a finite volume, similar to a discrete particle like a proton or neutron. Solitons occur only in nonlinear systems, since nonlinear effects are needed to cancel the natural spreading of the waves as they propagate \cite{agrawal, simon_top}.
\\
Vector solitons can be bright solitary waves that consist of localized high intensity pulses propagating through a darker background. They can also be dark soliton or pulses of darkness in a bright light field. They can be knotted and move through space with constant velocity as discussed by Nye \& Berry, in their seminal 1974 paper in which, they observed that wavefronts can contain dislocation lines closely resembling those  found in crystals \cite{nye}. The wavefront dislocations can be created as a loop or in pairs. The morphology of the dislocations was analyzed; they showed that the dislocations can be curved, they may intersect, collide, and rebound. Experimental studies in optics of waves with screw dislocations were later reported \cite{bazh, basis}; where a double screw dislocation or phase singularity was created in the process of diffraction of a plane wave or a gaussian beam by computer--synthesized gratings. \\
The mathematical expression for a wave with phase singularities are exact solutions of the Helmholtz equation. These wavefront dislocations are `cylindrical tubes' in which, the wave intensity vanishes and around which the phase changes by multiples of 2$\pi$. It is proposed that  a superposition of Bessel beams are ideal for the production of such singularities \cite{berry}. \\
 The ideal Bessel beams are also solutions to the Helmholtz equation. They are  immune to diffractive spreading and can heal themselves after being disrupted by an obstacle. Bessel beams as proposed by Durnin, have a well defined bright central spot radius surrounded by concentric rings  \cite{Durnin1987, Durnin1987exp} and the beam's intensity distribution is invariant along its propagation direction. Such idealized beam are  rigorously exact solutions, in infinite free space, of the scalar Helmholtz monochromatic wave equation. Any realizations of such beams in the laboratory   will require an infinite amount of energy. However, over a limited spatial range, approximations to the ideal Bessel beam can be obtained. One such light field is termed Bessel--Gauss (BG) beams \cite{simon}. \\
BG beams have an additional Gaussian factor  that decreases the light amplitude away from their center. They can be realized experimentally using conical lenses such as axicons \cite{Indenbetouw1989}, holographic masks \cite{Vasara1989}, or by spatial light modulators (SLM) \cite{SLM2017}.
 BG beams still exhibit the self-healing property \cite{Bouchal1998} in addition to the relatively diffraction-free propagation of ideal Bessel beams. These qualities have led to their adoption in a wide-range of fields which include, but are not limited to, optical communication\cite{Martelli2010, Gatto2011}, and various other applications as listed in \cite{McCloin2005}. 
 \\
 Nonlinear wave mixing is a method of interest for the superposition of BG beams.  It was proposed that the structure of the screw dislocations of different order Bessel-like beam can be obtained by nonlinear processes of Second Harmonic Generation (SHG) \cite{berry, bazh, basis}.  \\
SHG with laboratory-generated Bessel beams, is an area of great interest \cite{Arlt1999, laurell}. These experimental studies initially took place because it was thought that the non-diffracting region of high intensity would enable high conversion efficiencies \cite{shino}. At the same time, it was demonstrated that SHG excited by a Bessel-like distribution is gradually changed into a high-brightness axial frequency doubled beam by adjustment of the phase-matching conditions in the crystal \cite{laurell}.

Here, we experimentally show for the first time, that the second harmonic (SH) light generated by Bessel--Gauss beams are vector solitons knotted by ellipsoidal concentric rings. We show that their spatial profile is invariant against propagation. We observe that their behavior is similar to that of screw dislocation in solitary wave to resemble a soliton-like scattering: they collide and rebound at a 90$^\circ$ angle but retain their identity as they propagate.
\\
The BG beams are constructed by means of  virtual axicons encoded as a hologram on an SLM. We closely follow the methods detailed in Rosales--Guzm\'an and Forbes \cite{SLM2017}. We demonstrate experimentally that by increasing the topological charge ($\ell$) of the Bessel function, that produces the  BG beam, we can generate a frequency doubled light with spatial properties that closely resemble that  of  screw dislocations \cite{holler}. The zeroth order ($\ell = 0$)  BG beam produces two bight central solitons linked and knotted together by indistinguishable vortex lines. The higher-order BG beams generate coupled Bessel rings with phase singularities where the light intensities completely vanish.
All vector soliton geometries are surrounded by concentric rings with decaying brightness away from their central core; similar to that of a BG beam. The coupled solitons retain the self--healing properties of the BG beam by rotating and adopting a spring-like oscillation as they propagate through space. The ellipsoidal structure of the generated beams determines the global topology of the field with which the loop could be threaded. The superposition of the BG beams allows for the second harmonic light to form  links that are threaded and knotted,  enabling the solitons to intersect or collide, rotate, and rebound. 

\section{Preliminary}
 The ideal Bessel beams have electric field amplitude described mathematically by :
 \begin{equation}
\label{ideal_bessel}
\vb{\mathrm{E}}(r, \varphi, z, t) =  \mathcal{E} e^{i k_z z} J_{\ell}(k_r r)\times e^{i\ell\varphi}
\end{equation}
where $\mathcal{E} $ is the amplitude, $J_{\ell}(k_r r)$ are the Bessel functions of order $\ell$, while $k_r$ and $k_z$ are the radial and longitudinal components of the wave vector  $\va{k}$, respectively, obeying the relation $\abs{k}^2 = k_r^2 + k_z^2$. \\
This beam has an infinite extent and its intensity profile is a bright central core surrounded by concentric rings. Each ring carries approximately the same power as the central spot. The beam's intensity profile is also propagation invariant. This combination makes it impossible to realize experimentally. There are, however, a number of ways to generate good approximations. \\
Here, we use the Bessel-Gauss approximation where an additional Gaussian factor is introduced that decreases the amplitude away from the origin ($z = 0$), as follow \cite{SLM2017}:
 \begin{equation}
\label{bessel_gauss}
\vb{\mathrm{E}}(r, \varphi) =  \mathcal{E} e^{i k_z z} J_{\ell}(k_r r)\times e^{i\ell\varphi}\times e^{-\qty(\frac{r}{w_o})^2}
\end{equation}
where $w_o$ is a measure of the input Gaussian beam waist and is related to the BG  finite propagation distance $z_{max}$ as: $\frac{z_{max}}{w_o}=  \frac{k}{k_r}$. \\
We generate the BG beams  with an SLM by encoding an azimuthal variation (of modulus $2\pi$) and a blazed grating to separate the first order from the others \cite{SLM2017}. The mathematical expression has the following form:  
\begin{equation}
	\label{eqn:mask}
          \Phi_{SLM}=\mod \qty[k \alpha (n-1) r + \ell \phi 
                               + 2\pi(G_{x}x+G_{y}y), 2\pi]
\end{equation}
where `mod' is the modulus function, and $G_x$ and $G_y$ are the grating frequencies along the $x$ and $y$ directions, $n$ is the refractive index of the axicon, $\alpha$ is the angle of the axicon: the apex angle of the conical prism,  $\ell$ is the so-called topological charge or azimuthal order of the BG beam.
\section{Experiment}
\begin{figure}[h]
\centering
\includegraphics[width=0.46\textwidth]{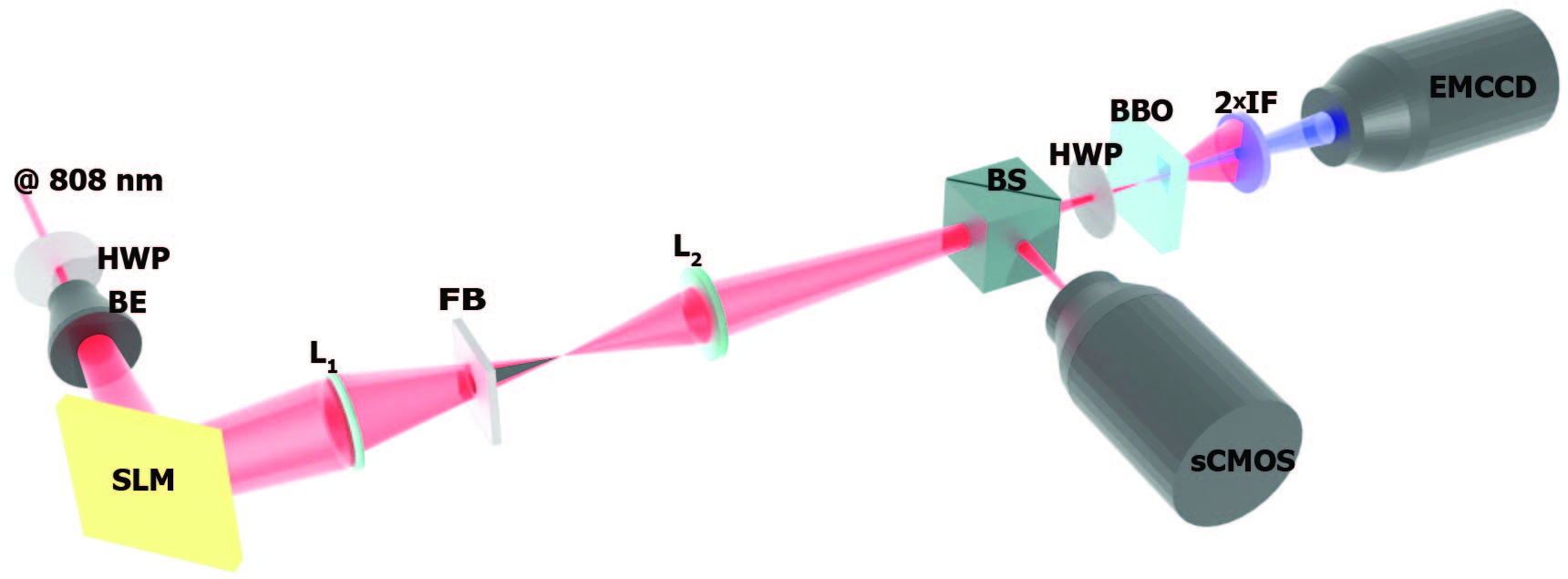}
\caption{Schematic of the experimental setup:  the signal from the laser is collimated by the beam expander (BE) to  cover the SLM. The structured beam from the SLM hologram is collected by the lens L$_1$ and isolated from the remaining gaussian by the fundamental block (FB). The polarization of the fundamental is controlled by the half wave plate (HWP) before it is focused onto the BBO crystal by L$_2$. The pump signal is characterized using the sCMOS before it is blocked by the filters (IF) and the SHG signal is characterized using the EMCCD}
\label{fig:Setup}
\end{figure}

An illustration of the experimental setup is shown in FIG. \ref{fig:Setup}. The laser source is a mode locked regenerative Ti:sapphire laser with repetition rate of 3 kHz,  pulse energy of 1.67 mJ. The laser system can produce  a 38 fs  pulse duration at 808 nm center wavelength. The linearly polarized output of the laser is expanded to cover the  face of the light modulator. The wavefront manipulator is  a phase-only Santec reflective liquid crystal on silicon (LCOS ) based Spatial Light Modulator (SLM--200). The nonlinear medium we use for frequency doubling, is  a 500 $\mu$m thick type-I $\beta$--Barium Borate (BBO) crystal, with optical axis $\theta = 28 ^{\circ}$. The crystal  is affixed to a kinematic mount which allows translation along the optical axis, and to be tilted and rotated for optimal phase matching. We generate the desired phase mask or holographic grating on a computer to display on the SLM.
\\ 
All the diffracted orders generated  from the SLM masks are collected by a lens L$_1$ and focused through the BBO by L$_2$. We have designed a homemade aperture (FB) to block the remaining gaussian beam. 
The sCMOS camera is used to characterize the spatial distribution and intensity of the beam incident on the BBO. The sCMOS and  BBO crystal are equidistant from the  beamsplitter (BS) to ensure that the desired beam mode is incident on the crystal. A second HWP is placed before the BBO to further optimize the phase matching conditions. 
The generated second harmonic signal is imaged using an EMCCD placed in the far-field. We use two interference filters centered at 405 $\pm$ 5 $nm$ to ensure that only the frequency doubled signals (SH) are imaged.
\\
Throughout the experiment, we make full use of the flexibility that Eq. (\ref{eqn:mask}) allows. We are able to control the generated hologram  using two different methods:
 \\
 (i)  In this first case, we keep the topological charge $\ell$ of the BG beam constant and set to zero. We vary the refractive index, which changes the depth-of-field (DOF) of the virtual axicon. The DOF  is a function of the radius ($w_o$) of the beam `entering'  the axicon (reflected off the SLM), the axicon's index of refraction ($n$), and its opening angle ($\alpha$); $w_o$ and $\alpha$ are kept constant throughout. The $DOF$ is given by: 
\begin{equation}
\label{DOF}
DOF \approx \frac{R}{(n - 1)\alpha}
\end{equation} 
where the virtual axicon is defines by its alpha ($\alpha$) and apex angle, $R$ is the radius of the beam incident on the SLM.
 \\
(ii) We can also control the encoded holograms by keeping the refractive index of the axicon constant at $n = 1.5$ and varying the topological charge $\ell$ of the BG beam. 

\section{Results and Discussion}
\begin{figure}[h]
	\centering
	\includegraphics[width=0.46\textwidth]{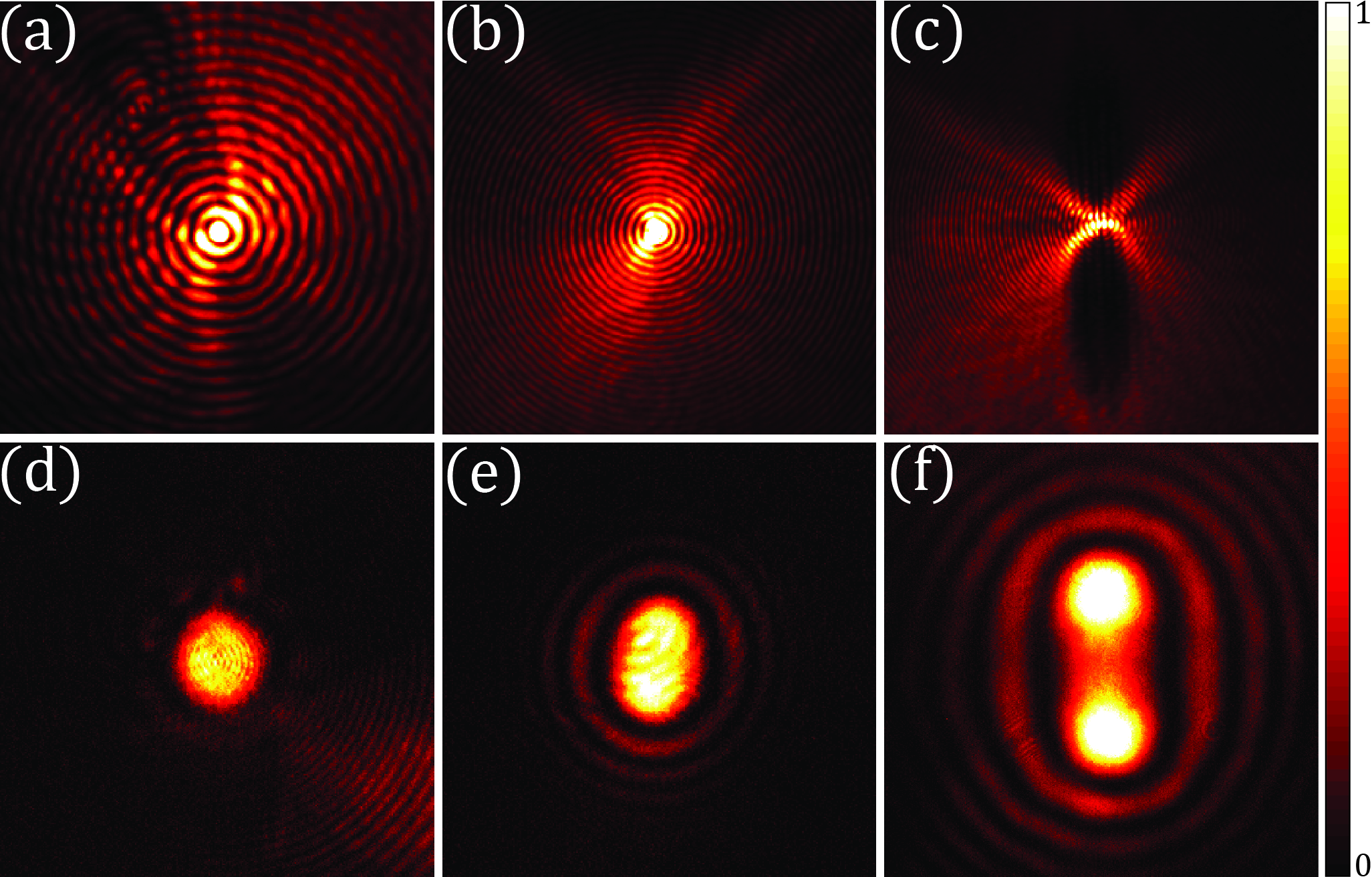}
	\caption{(a)--(c) are the intensity profile of the BG beams incident on the BBO. They are generated by the SLM and imaged using the sCMOS. (d)--(f) are the corresponding intensity distribution of the SH light generated; imaged using the EMCCD. When the virtual axicon refractive index is $n=1.5$ the input beam is shown in (a), which generates the output in (d). For  $n=2.0$, its input beam is (b) and its output is (e) showing that its central core has an ellipsoidal intensity distribution. For $n=2.5$ the  input BG beam incident on the crystal is displayed in (c) and its corresponding SH output is (f) composed of two linked solitons.}
	\label{fig:SHG}
\end{figure}

\begin{figure}
\begin{center}
\includegraphics[width=0.46\textwidth]{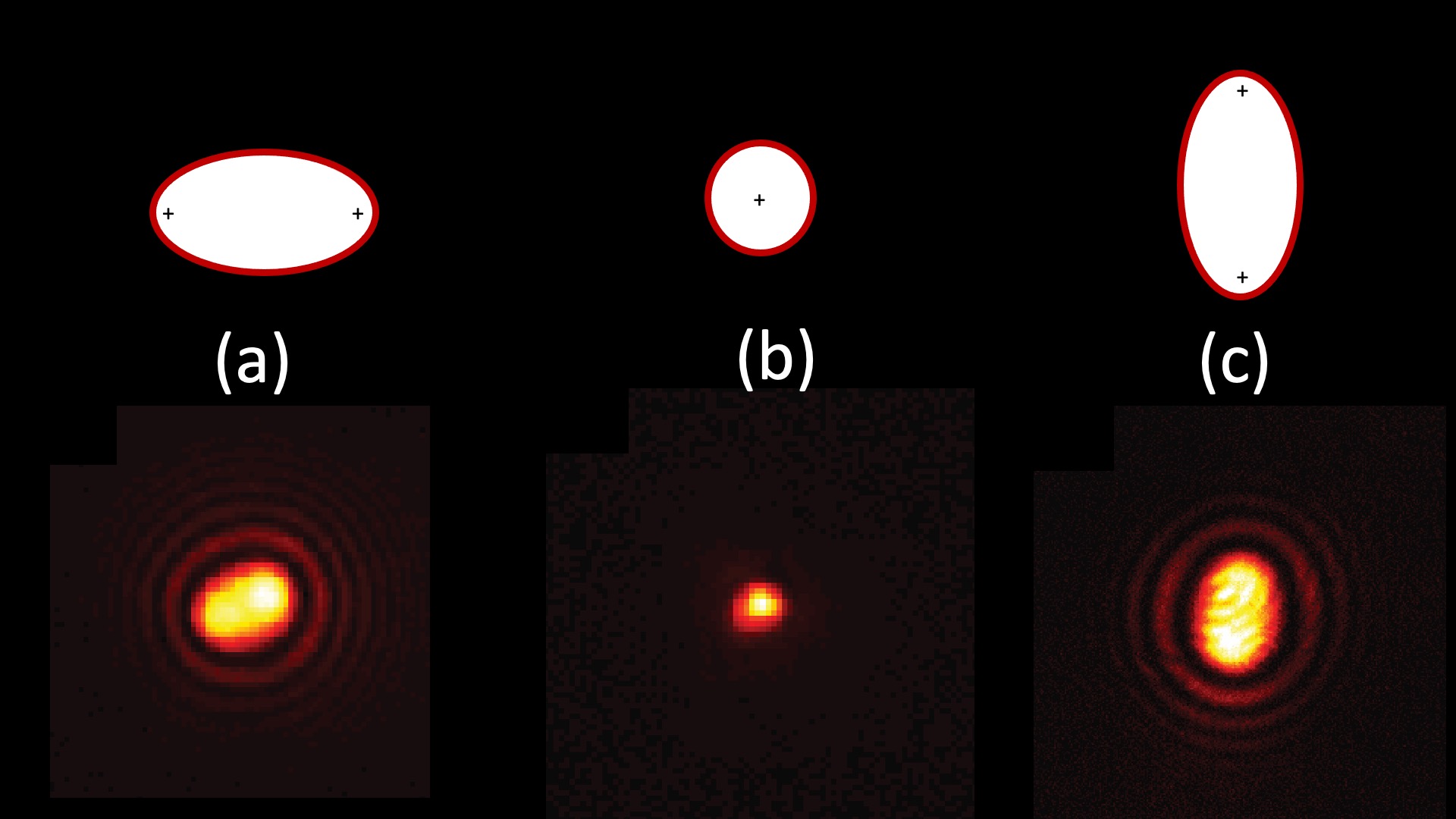}
\caption{The evolution of the  SH ellipsoidal BG beam generated in FIG \ref{fig:SHG}--(e) is shown. (a)--(c) show that such an elliptical beam propagates and mirrors the smooth evolution of an ellipse and the right-angle scattering of its foci, as recorded \href{https://www.youtube.com/watch?v=UQUAjNP2zos}{here (Visualization 1)}}
\label{ellipse}
\end{center}
\end{figure}

\begin{figure}[h]
	\centering
	\includegraphics[width=0.46\textwidth]{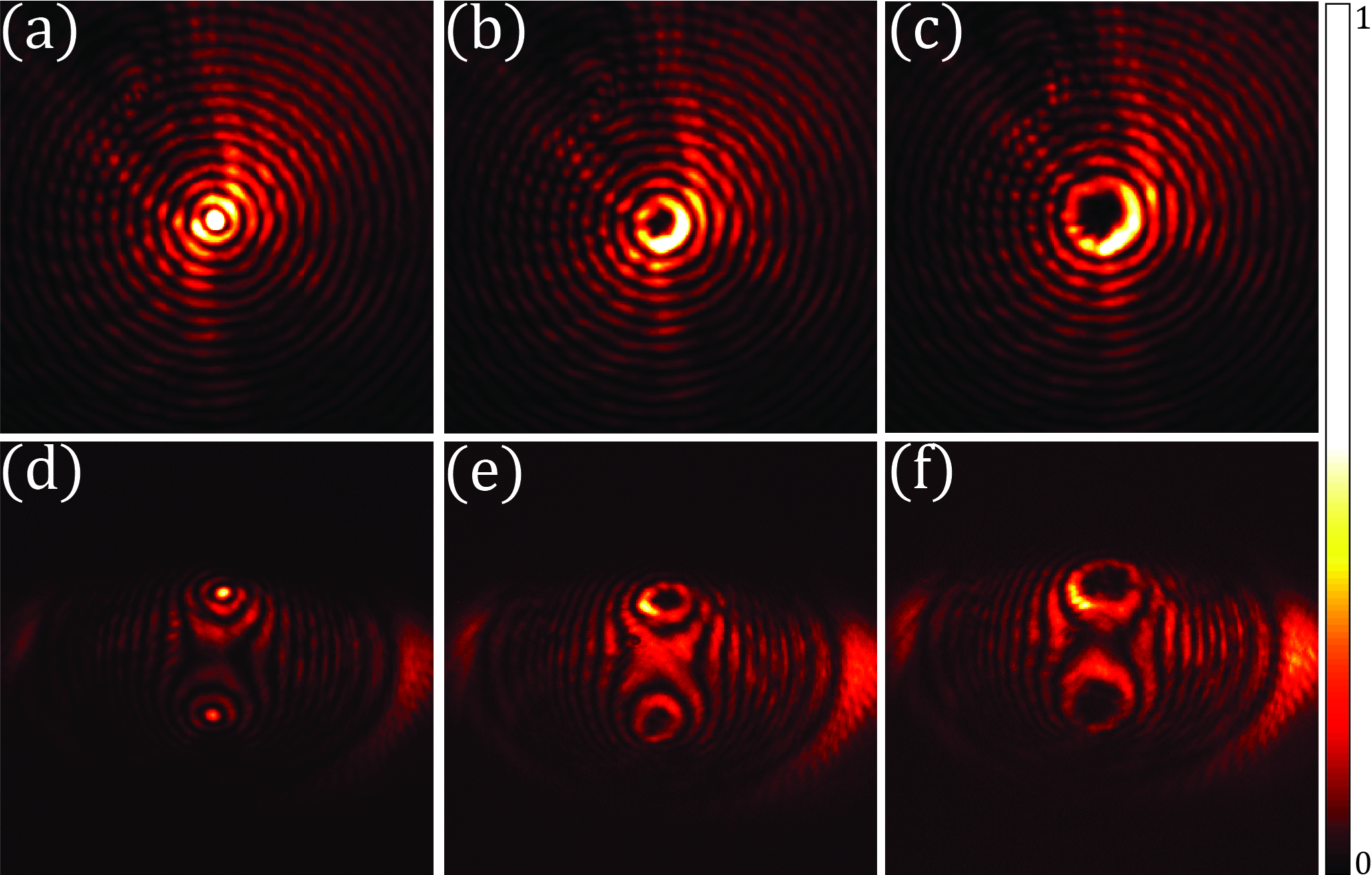}
	\caption{The top row are experimentally generated Bessel--Gauss beams; (a) $\ell = 0$, (b)  $\ell = 3$; and (c) $\ell = 5$. In the bottom row are the corresponding topological vector solitons: (d) the vector soliton are two coupled $\ell = 0$ BG beam, (e) the vector soliton are two $\ell = 3$ beam, and in (f) the vector solitons are two $\ell = 5$ beams.}
	\label{bessels}
\end{figure}

We start with method (i): in the first row of FIG. \ref{fig:SHG} we show the intensity distribution of the BG beam that is incident on the BBO crystal. One can observe that as the refractive index of the axicon is changed, the DOF of the virtual lens also varies as in equation \ref{DOF}.
\\
FIG. \ref{fig:SHG}--(a) shows the zeroth order BG beam for $n = 1.5$ as the fundamental input to the frequency doubled beam in FIG. \ref{fig:SHG}--(d). The resulting beam in FIG. \ref{fig:SHG}--(d) shows that  a considerable part of the SH signal is generated in the axial direction, along the axis of the BG cone as previously  demonstrated for a non-collinear phase matching, where wave vectors with the opposite radial components generate second-harmonic output along the beam axis and give a single on-axis output in the far field \cite{Arlt1999, laurell}. In addition to our experimental verification, we also observe that the single on axis beam is contained or bounded by  concentric Bessel rings.
\\
We then increase the refractive index to $n= 2.0$, FIG. \ref{fig:SHG}--(b) shows that the fundamental beam still maintains its zeroth order BG beam but the interference pattern at the DOF of the virtual axicon is quite noticeable. The resulting SH of such an input beam is shown in FIG. \ref{fig:SHG}--(e). The frequency doubled beam still displays  a considerable part of its intensity in the axial direction, along the axis of the BG cone. However, this particular on-axis beam has an ellipsoidal shape contained or bounded by ellipsoidal rings or vortices. The bounding Bessel rings allow for the ellipsoidal central spot of the frequency doubled beam to retain the non-diffracting properties of the fundamental Bessel beam. As far as the authors know, this is the first observation of an ellipsoidal soliton generated through SHG of BG beams. 
\\
We next set the refractive index of the virtual axicon to $n = 2.5$, the incident light beam in FIG. \ref{fig:SHG}--(c) has intensity distribution similar to that found within the DOF of the axicon. In this way, we allow the zeroth order beam generated by this virtual axicon to form within the bulk of the nonlinear crystal. FIG. \ref{fig:SHG}--(f) shows the resulting SH light generated. It is composed of two central spots (Gaussian--dots: GD) of similar radius that are linked and knotted by ellipsoidal concentric rings. The GD in our experimental finding have properties similar to the topological optical solitons or ``\textit{knotted nothings}'' observed in \cite{bazh, basis} and  discussed in Berry \textit {et al.}\cite{berry}. \textit{Knotted nothings}  are wavefront dislocations or phase singularities where the wave function vanishes. In our findings, instead of singularities, we observe bright center spots or GD, where the light intensity is maximum. 
 Such composite solitons have topologies dependent on the topological charge $\ell$ of the fundamental BG beam used to create them. The coupled vector solitons in FIG \ref{fig:SHG}--(f) is an experimental demonstration of coupled vortex vector solitons \cite{agrawal, muslim}. They are a superposition of two zeroth order BG beams, that through the nonlinear mixing process, have their bright central spots linked and knotted together by two indistinguishable vortex lines. These topological vector solitons were first suggested theoretically by Manakov \cite{manakov}, assuming the self-- and the cross--phase modulations are equal in the nonlinear wave mixing process.
 \\
To demonstrate its solitary wave properties, we let the ellipsoidal-beam in FIG. \ref{fig:SHG}--(e) propagate for approximately a 1.5 meters in the lab. In FIG. \ref{ellipse}, we observe that as the beam propagates it displays the right-angle scattering phenomenon observed during the right-angle scattering of two vortices in a head-on collision \cite{manton}. 
The ellipse-beam propagation is analogous with the evolution of an ellipse as the parameters change.
Consider the equation,
\begin{equation}
\label{ellip}
\frac{x^2}{(1 + \epsilon)^2} + \frac{y^2}{(1 + \epsilon)^2} = 1
\end{equation}
as $\epsilon$ moves through zero. This defines an ellipse whose shape for $\epsilon < 0$ in FIG. \ref{ellipse}--(a) , $\epsilon = 0$ in FIG. \ref{ellipse}--(b), and $\epsilon > 0$ is shown in FIG. \ref{ellipse}--(c).
The foci are at $\qty(\pm 2\sqrt{\abs{\epsilon}}, 0)$ for $\epsilon \leq 0$ and at $\qty(0, \pm 2\sqrt{\epsilon})$ for $\epsilon \geq 0$. They scatter through a right angle while the ellipse smoothly changes shape. The experimental ellipsoidal soliton beam in FIG. \ref{fig:SHG}--(e) displays this property as it propagates; this  is recorded {\href{https://www.youtube.com/watch?v=UQUAjNP2zos}{\texttt{here}}}.
 \\
To further investigate the geometric properties and spatial distribution of the vector solitons, we now vary the topological charge $\ell$ for the fundamental input BG beam while maintaining the refractive index of the virtual axicon at $n = 1.5$ in equation \ref{eqn:mask}, as detailed in method (ii) above. 
\\
In this way, we show that higher-order BG beam can also form knotted vector solitons through the frequency doubling process; as displayed in FIG. \ref{bessels}.  The top row are the fundamental input BG beams and the bottom row are their corresponding SH output. In FIG. \ref{bessels}--(d) we generate a more complex vector solitons where the central spot of the $\ell = 0$ BG beam together with the 2 first bessel rings are knotted together by a larger spheroidal Bessel ring. This is obtained by varying the angle of incidence between the fundament input beam and the nonlinear crystal. Figures \ref{bessels}--(e) and (f) are the vector solitons obtained with BG beam with topological charge $\ell = 3$ and  $\ell = 5$ respectively. Here again the 
higher-order BG beams are knotted together through the SHG process. The higher topological charge beams have phase singularities observed in screw dislocation at which, the amplitude of the electric field becomes zero at the center of the beam\cite{nye, bazh, basis}. In this highly nonlinear process, momentum conservation leads to phase matching conditions so that two photons at the  same frequency add to produce one photon with twice that frequency. Similarly, the fundamental beam with a phase singularity transforms into a beam with two phase singularities. Associated with the geometry and size of the singularities is the topological charge of the input Bessel beam.
\\ 
Probably the most interesting phenomenon in the vector soliton dynamics is that the intensity structure of the solitons
constituents rotates throughout propagation. This behavior is similar to two-vortex right-angle scattering that occurs in a head-on collision after which, the knotted solitons are classically indistinguishable and can not be labelled.
The soliton dynamic we observe follow the numerical model of nonlinear evolution equations developed by S. J. Orfanidis \cite{Orfanidis80}, where  the solitons solutions of the O(3) $\sigma -$ model are considered in two Euclidean (2 + 1) dimensions, and their scattering properties are predicted \cite{Leese90, Leese900, Zakrzewski91}.
The simplest Lorentz-invariant O(3) model in ($2+1$) dimension is defined by the Lagrangian density:
\begin{equation}
\mathscr{L} = \frac{1}{4}(\partial^\mu \vb*{\phi})\vdot(\partial_\mu \vb*{\phi})
\label{eqn:Ld}
\end{equation}
 which contains three real scalar fields, $\vb*{\phi} \equiv$ ($\phi_1, \phi_2, \phi_3$). In ($2+1$) dimensions, $\vb*{\phi}$ is a function of the spacetime coordinates ($t, x , y$), with the constraint that  $\vb*{\phi}\vdot \vb*{\phi^*} = 1$.
 \\
 The formulation we adopt for our model, use one independent complex field $W$, such that 
 \begin{equation}
\begin{split}
\phi_1 & = \frac{W+W^*}{ 1 +|W|^2} \\
\phi_2 & = i\frac{W-W^*}{ 1 +|W|^2} \\
\phi_3 & = \frac{1-|W|^2}{ 1 +|W|^2}\\
\end{split}
\label{eqn:5}
\end{equation}
 and the Lagrangian density is given by:
  \begin{equation}
\mathscr{L} = \frac{\partial^\mu W \partial_\mu W^*}{(1 + |W|^2)^2} 
 \label{eqn:6}
\end{equation}
The general two static solitons configuration corresponds to:
\begin{equation}
\label{eqn:7}
W = \lambda \frac{(x + iy - a)(x + iy - c)}{(x + iy - b)}
\end{equation}
where $a$, $b$, $c$, $d$ and $\lambda$ are arbitrary complex numbers. The positions and sizes of the static soliton can be determined in terms of these complex numbers. In $(2 + 1)$ dimensions the solitons may be made to move with arbitrary velocity: the positions of solitons are given by $\pm a$ (we choose $a = c$) so that the solitons widths correspond $2\lambda a (\abs{a \pm b})^{-1}$. We then 
 chose to modify $ a \rightarrow a' = a(1 - Vt)$, $V$ is the velocity of the soliton. A similar transformation is also made for $b$ as this expression corresponds to solitons moving towards each other while not changing their size \cite{Zakrzewski91}.
\\
We next insert the modified expressions for $a$ and $b$ into $W$ in equation \ref{eqn:7}. We can now calculate $\partial_t W$ and use the calculated values of $W$ and $\partial_t  W$ (at $t = 0$) as the initial values for our simulations.
\begin{figure*}
	\centering
	\includegraphics[width=\textwidth]{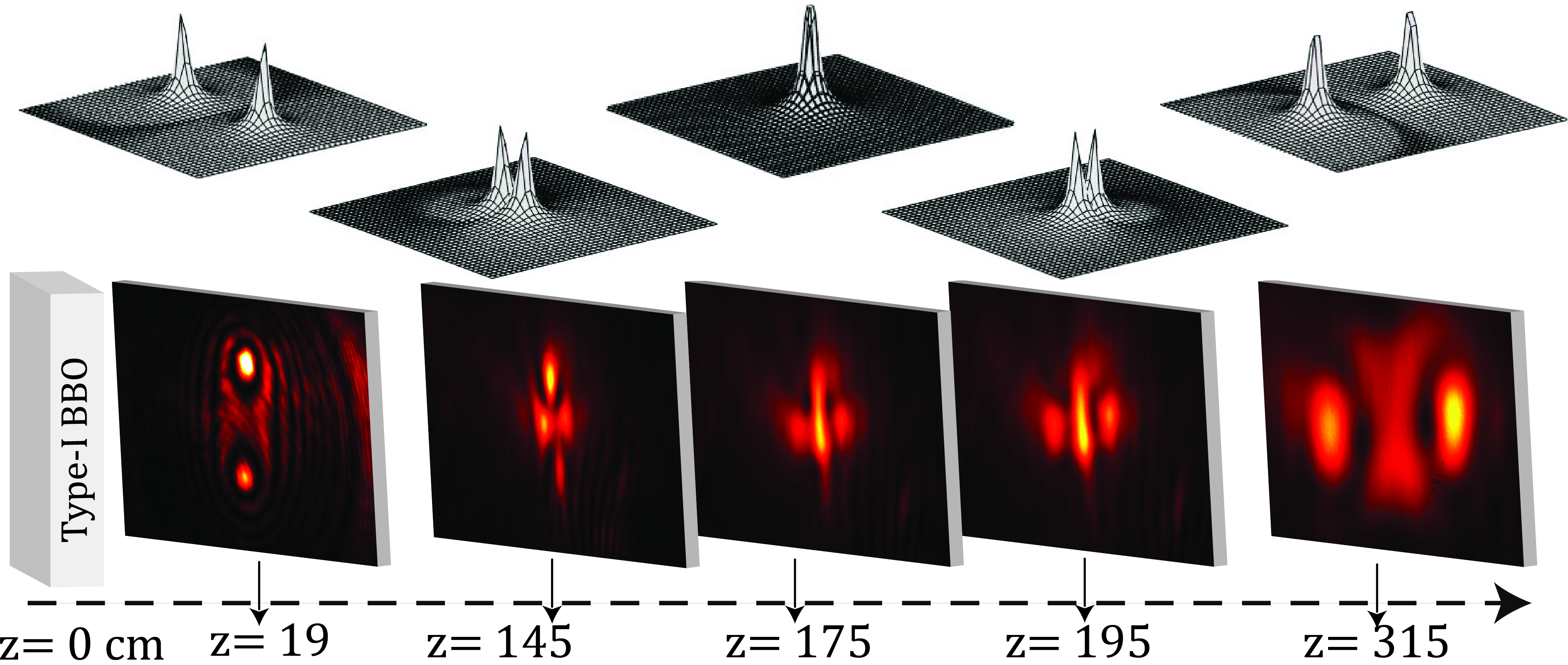}
	\caption{Comparison of energy density plots at increasing times during the right-angle scattering of two vector solitons in a head-on collision; $W$ and $\partial_t  W$ are given by equation \ref{eqn:7}. A full recording of the solitons evolution can be viewed \href{https://youtu.be/SjgTa_-9QHk}{here (Visualization 2)}.}
	\label{transition}
\end{figure*}
\\
In FIG. \ref{transition} we compare our numerical simulations to our experimental data. We can clearly observe that the two solitons scattering resembles a head-on collision; that is the solitons come together, form a complicated structure then they separate at an angle close to $90^\circ$ \href{https://youtu.be/SjgTa_-9QHk}{ (Visualization 2)}. 
\\
To capture the vector soliton dynamics in the laboratory, we first place the detector 20 cm away from the nonlinear crystal. We then translate the camera  and take pictures at fixed intervals to capture the scattering behavior of the coupled solitons. We can clearly see at FIG. \ref{transition} at $ z = 175$ \& $195$ cm that the solitons appear to shrink and morphed together before separating again at an angle perpendicular to the direction of travel. After the solitons have separated their trajectories are again well defined and the continuation suggests that the motions is as shown in the numerical simulation. Interestingly, the two solitons can be followed 
 until collision, and treated as particle trajectories, but when the solitons emerge at right angles, one can no longer say which outgoing solitons corresponds to which incoming one. One interpretation of what happens in the collision is that
some of the energy making up each soliton is exchanged. Each outgoing soliton is made up of one half of each ingoing solitons. In such a case, the vector solitons become indistinguishable and cannot be treated with a solely classical approach.

\section{Conclusion}
In our experiment, we have observed for the first time to our knowledge, that the second harmonic signal from Bessel-Gauss beams with various topological charge, are coupled vector solitons that are knotted and linked by ellipsoidal Bessel rings.  The BG beams with higher-order topological charge transform into screw dislocation with phase singularities through the process of second harmonic generation.
We further show that these vector solitons collide, rebound, and form complicated structures as they propagate, before separating at a  $90^\circ$ from their direction of travel. We also observe that the SH beams maintain some of the  non-diffractive properties inherited from the BG beams used to generate them. 
Further studies of our results, specifically the two-dot configuration in FIG. \ref{fig:SHG}--(f), could be used to describe stationary configurations of electrons in free space where Schr\"odinger's equation reduces to the Helmholtz equation. One can also note that as the GD collide and flip direction, the two become indistinguishable. For this reason we believe that our experimental findings could spur new methods of generating entangled photons.
Further, our experimental finding  can have topological textures, analogous to textures in condensed matter and high-energy physics, to name a few. This could  lead to  further insights and possibilities for topologically structured light and its applications.
\begin{acknowledgments}
The authors wish to thank Professors Herbert Winful, Kimani Toussaint, Theodore Norris, and Thomas Searles for their time, conversations, advice, and research support.
\\
This research is supported in part by the National Geospatial Intelligent Agency grant \# HM04762010012.
\end{acknowledgments}
\newpage

\end{document}